\documentclass[a4paper,11pt]{article}
\pdfoutput=1 

\usepackage{jinstpub} 

\usepackage[utf8]{inputenc}
\usepackage{circuitikz}
\usepackage{multirow}
\usepackage{booktabs}
\usepackage{comment}
\usepackage{float}

\title{\boldmath Characterization of photomultiplier tubes in a novel operation mode for Secondary Emission Ionization Calorimetry}

\author[a,1]{E. Tiras,\note{Corresponding author.}}
\author[a]{K. Dilsiz,}
\author[a]{H. Ogul,}
\author[a]{D. Southwick,}
\author[a]{B. Bilki,}
\author[a]{J. Wetzel,}
\author[a]{J. Nachtman,}
\author[a]{Y. Onel,}
\author[b]{D. Winn}

\affiliation[a]{Department of Physics and Astronomy, University of Iowa\\203 Van Allen Hall, Iowa City, IA 52242, USA}
\affiliation[b]{Department of Physics, Fairfield University\\BNW118, 1073 N. Benson Rd., Fairfield, CT 06824, USA }

\emailAdd{emrah-tiras@uiowa.edu}

\abstract{Hamamatsu single anode R7761 and multi-anode R5900-00-M16 Photomultiplier Tubes have been characterized for use in a Secondary Emission (SE) Ionization Calorimetry study. SE Ionization Calorimetry is a novel technique to measure electromagnetic shower particles in extreme radiation environments. The different operation modes used in these tests were developed by modifying the conventional PMT bias circuit. These modifications were simple changes to the arrangement of the voltage dividers of the baseboard circuits. The PMTs with modified bases, referred to as operating in SE mode, are used as an SE detector module in an SE calorimeter prototype, and placed between absorber materials (Fe, Cu, Pb, W, etc.). Here, the technical design of different operation modes, as well as the characterization measurements of both SE modes and the conventional PMT mode are reported.}

\keywords{Calorimeters, Detector modelling and simulations II (electric fields, charge transport, multiplication and induction, pulse formation, electron emission, etc), Photon detectors for UV, visible and IR photons (vacuum) (photomultipliers, HPDs, others), Modular Electronics}

\arxivnumber{1605.00667}

\begin{document}
\maketitle
\flushbottom

\section{Introduction}
\label{sec:intro}
The increasing instantaneous luminosity and resulting unprecedented radiation conditions at particle accelerators, such as the Large Hadron Collider (LHC), necessitate a major detector transformation in order to maintain/improve physics measurement capabilities. These highly demanding experimental conditions require radiation-hard detectors and calorimeters for robust and reliable performance. In this context, we have studied Secondary Emission \cite{PhysRev.107.977} (SE) Ionization Calorimetry as a new, novel particle calorimetry technique in high radiation environments. Since there are no dedicated secondary emission modules commercially available, we developed adaptations in the biasing modes of photomultiplier tubes (PMTs) to demonstrate the secondary emission mechanism in the dynode chain and the possibility to use such detectors in calorimetry. 

Traditional PMT baseboards create a high potential difference between the photocathode of the PMT (usually coated metals with a favorable work function) and the anode of the PMT, with several intermediate dynodes at increasing potentials in between.  A photon striking the photocathode induces electron emission. Those electrons are then accelerated through the potential difference directly into the next PMT dynode.  This knocks more free electrons, which are accelerated into the next dynode, and so forth, creating a cascading amplifying effect. This amplification process provides the experimenter with sufficient signal to detect a single photon.

In an SE module, Secondary Emission electrons (SEe) are generated from an SE surface, cathode or dynodes, when charged particles (shower particles) penetrate the module \citep{winnonel, snowmass}. An SE cathode is a thin film. These films are typically simple metal-oxides Al$_2$O$_3$, MgO, CuO/BeO, or other higher yield materials. These materials are known to be very radiation-hard, as they are used in PMTs (up to 50 Grad dose) and in accelerator beam monitors (flux of >1020 MIP/cm$^{2}$).

On the inner surface of a metal plate in vacuum, which serves as the entrance "window" to a compact vacuum vessel (metal or metal-ceramic), an SE film cathode is analogous to a photocathode, and the shower particles are similar to incident photons. The SEe produced from the top SE surface by the passage of shower particles, as well as the SEe produced from the passage of the shower particles through the dynodes, are similar to photoelectrons. The statistics of photoelectrons and SEe are similar. The SEe are then amplified by sheets of dynodes.

By observing the intrinsic similarities of the PMTs and the envisaged SE modules, a PMT can be used to validate the concept of an SE module with simple modifications. In order to validate this approach, we started with conventional PMTs. The PMTs were selected such that they had excessive usage, hence had potentially degraded photocathode performance. Although it would be ideal to be able to use the photocathode as an SE film cathode mentioned above, the contributions from the \v{C}erenkov photons created at the glass entry window of the PMT cannot be prevented \cite{CMSHCAL} and would bias the SE measurements. Therefore, the validation procedure consists of disabling the photocathode from the multiplication chain rendering the PMT non-responsive to photons incident on the photocathode. In this case, the entire dynode chain is utilized as SE surfaces. The largest contribution comes from an initial SEe production at the first dynode. 

In order to study the effects of multiple schemes in disabling the photocathode, we developed two modules. The first module was constructed with seven Hamamatsu single anode R7761 and the second one with nine Hamamatsu 16 pixel R5900-00-M16 multi-anode photomultipliers \cite{R7761}. For the former, two different SE modes are developed in order to observe the effect of secondary emissions. One mode is the so-called `cathode float' (CF) mode, where the photocathode is removed from the signal chain.  The other mode is the so-called `cathode/first dynode shorted' (C-D1) mode, where the photocathode is included in the signal chain. For the multi-anode tubes, we have only developed the C-D1 mode. This is due to the fact that the multi-anode PMTs have metal channel dynode structure which has a significantly smaller fraction of the area of the photocathode. By applying C-D1 mode, cathode to D2 multiplication is enabled to some extent. 

We can also operate the PMTs in the traditional (PM) mode.  For the two different PMT types, two different circuit boards were developed to power them in all mentioned modes. Each PMT was tested for gain and dark current in both PM and SE modes. 

Here, we report on the technical design of SE operation modes and measurements performed on 74 PMTs. These tubes are used in a later calorimetric measurement study to be published as a separate article. 

\section{Technical Design}

The Hamamatsu single anode R7761 PMT has a bialakali photocathode material with borosilicate glass and it has 19 dynode stages that are secondary emissive electrodes in a fine mesh structure. The 16 anode R5900-00-M16 PMT has the same photocathode and window materials but metal channel dynode structure as the multiplication chain. The dynodes of both PMT types are either nickel, stainless steel or copper-beryllium alloy that are coated with different emissive materials such as BeO, MgO, GaP and GaAsP. Both PMT types were previously used for collecting data in the CDF experiment at Fermi National Accelerator Laboratory (Fermilab) \cite{Artikov, Grozis,Blair}.

\begin{figure}[h!]
\makebox[\textwidth]{%
\includegraphics[scale=0.22]{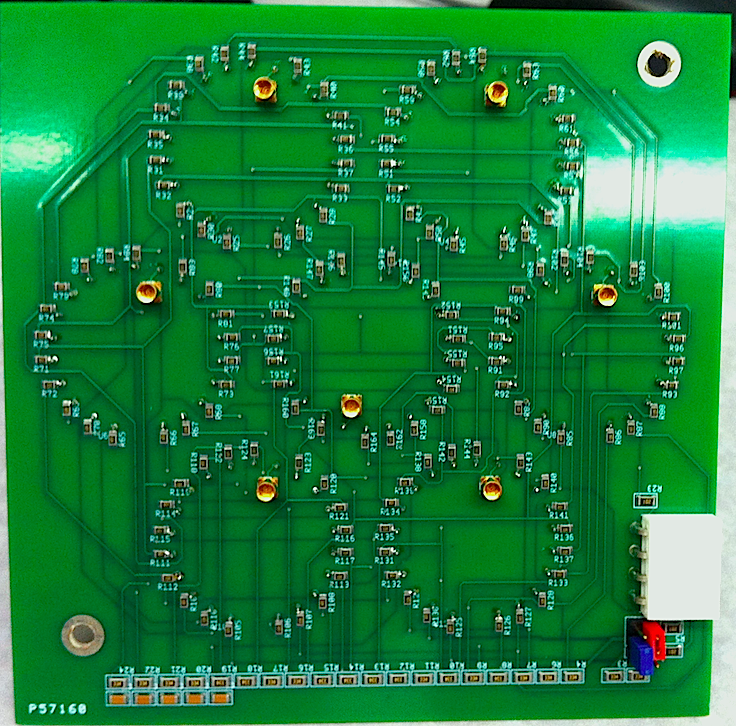}
\\
\includegraphics[scale=0.33]{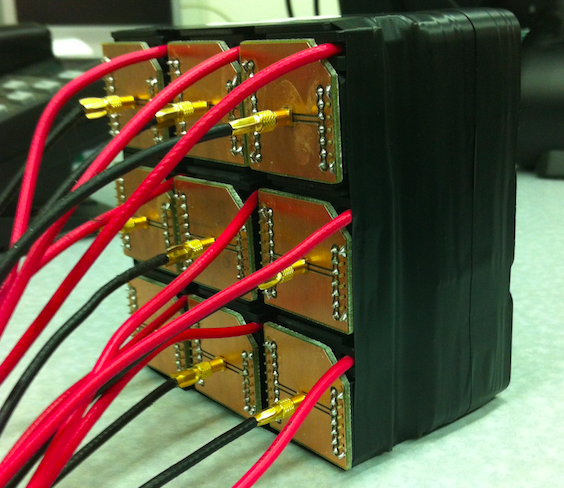}  
}%
\caption{Left: Circuit boards for powering seven R7761 PMTs, Right: Bases for powering each R5900-00-M16 PMTs. Both designs have switches/jumpers to select the mode of operation.}
\label{BaseboardforR7761-R5900} 
\end{figure}

A circuit board was designed to power seven Hamamatsu R7761 tubes with negative high voltage (Fig. \ref {BaseboardforR7761-R5900}  left). Each tube was powered by a common voltage divider located on the board. The values for this voltage divider are chosen following the reference design by Hamamatsu. 

Figure \ref {BaseboardforR7761-R5900} right shows the bases for powering each Hamamatsu R5900-00-M16 PMT. The assemblies shown in Fig. \ref {BaseboardforR7761-R5900} were primarily constructed for the SE calorimeter tests. 

Figure \ref{SE Diagram for R7761} shows the schematic of the voltage divider for the R7761 baseboard. The values of resistors R2-R20 are chosen such that they follow closely the reference design from Hamamatsu for negative high voltage operation, and provide a linear voltage divider. Despite depletion of the photocathode and considering the population of the board, base current is only slightly increased over the nominal single-socket reference current operating at about 300$\mu$A. To mitigate standing current depletion during pulses with multiple photomultiplier tubes, capacitors C1-C5 are added to store charge for the final dynodes. Finally, $R_i$ is used to attenuate noise introduced from the high voltage supply. 

Three different modes of operation exist on the baseboard for R7761 PMTs: 

\textbf{Mode 1 - normal divider or PM mode:} In this mode, the voltage divider chain is unmodified and the potential difference seen across the dynodes is equal, with the exception for 2x potential across the C-D1 gap. This is the reference design from Hamamatsu. 

\textbf{Mode 2 - cathode-first dynode (C-D1) shorted:} In this mode, jumpers on the board enable the bridging of R1, so that there is 0 potential across the C-D1 gap ($V_C$-$V_{D1}$ = 0V). 

\textbf{Mode 3 - cathode separated, cathode independent or cathode float mode:} The design of the board allows the cathode to be separated from the remainder of the divider chain. A second voltage input to the baseboard allows specific tuning of the voltage present on D1. With proper jumper selection and two high voltage sources, the potential across the C-D1 gap can be manipulated such that it becomes positive with respect to the gap of D1-D2. In this mode, if a second high voltage source is not used, there is a slight possibility that the photocathode can still be charging up. Previous tests with a single R7761 and a positive high voltage base applying +2kV to the anode and a few hundred volts to the cathode keeping D1 at ground were performed. The cathode positive voltage was varied slowly down to ground and there was no notable change in response to beam particles. In any case, Mode 3 is not the foreseen operation mode for a future, dedicated SE detector. It serves as a comparison tool in the framework of this study. 

All of these modes can be examined in Fig. \ref{SE Diagram for R7761}, where A-B bridge forms normal operation mode (Mode 1) with HV input on HV1, B-C bridge forms Mode 2 with HV input on HV1, and B-D bridge forms Mode 3 with HV input on HV2. 

\begin{figure}[h!]
  \begin{centering}
    \begin{circuitikz}[scale=0.4, transform shape]
      \draw (0,-4)node[anchor=west] {HV1} to[R=$R_i$,o-] (0,-1) to[short,-*] (0,5) node[anchor=west] {Cathode}; 
      \draw (0,2)to[R=$R_1$,*-] (2,2) to[short,-*] (2,-0.5)node[anchor=west] {A}; 
      \draw (0,2)--(0,0)to[C=$C_1$,*-*](2,0);
      \draw (0,-1)to[short,*-*](2,-1) node[anchor=west]{C};
      
      \draw (3,-4) node[anchor=west] {HV2} to[R=$R_i$,o-*] (3,-1)node[anchor=west] {D};
      \draw (3,-0.5) node[anchor=west] {B}to[short,*-*] (3,4) node[anchor=west] {D1};      
      \draw (3,2)to[R=$R_2$,*-*] (5,2) to[short,-*](5,4) node[anchor=west] {D2};
      \draw (3,2)--(3,0)to[C=$C_2$,*-*](5,0);
      \draw (5,2)to[R=$R_3$,-*] (7,2) to[short,-*](7,4) node[anchor=west] {D3}; 
      \draw (5,2)--(5,0)to[C=$C_3$,-*](7,0);
      \draw (7,2)to[R=$R_4$,-*] (9,2) to[short,-*](9,4) node[anchor=west] {D4};
      \draw (7,2)--(7,0)to[C=$C_4$,-*](9,0);
      \draw (9,2)to[R=$R_5$,-*] (11,2) to[short,-*](11,4) node[anchor=west] {D5};
      \draw (9,2)--(9,0)to[C=$C_5$,-*](11,0);
      \draw (11,2)to[R=$R_6$,-*] (13,2) to[short,-*](13,4) node[anchor=west] {D6};
      \draw (11,2)--(11,0)to[C=$C_6$,-*](13,0);
      \draw (13,2)to[R=$R_7$,-*] (15,2) to[short,-*](15,4) node[anchor=west] {D7};
      \draw (13,2)--(13,0)to[C=$C_7$,-*](15,0);
      \draw (15,2)to[R=$R_8$,-*] (17,2) to[short,-*](17,4) node[anchor=west] {D8};
      \draw (15,2)--(15,0)to[C=$C_8$,-*](17,0);
      \draw (17,2)to[R=$R_9$,-*] (19,2) to[short,-*](19,4) node[anchor=west] {D9};
      \draw (17,2)--(17,0)to[C=$C_9$,-*](19,0);
      \draw (19,2)to[R=$R_{10}$,-*] (21,2) to[short,-*](21,4) node[anchor=west] {D10};
      \draw (19,2)--(19,0)to[C=$C_{10}$,-*](21,0);
      \draw (21,2)to[R=$R_{11}$,-*] (23,2) to[short,-*](23,4) node[anchor=west] {D11};
      \draw (21,2)--(21,0)to[C=$C_{11}$,-*](23,0);
      \draw (23,2)to[R=$R_{12}$,-*] (25,2) to[short,-*](25,4) node[anchor=west] {D12};
      \draw (23,2)--(23,0)to[C=$C_{12}$,-*](25,0);
      \draw (25,2)to[R=$R_{13}$,-*] (27,2) to[short](27,4) to[short,-o] (28,4)node[anchor=west] {SIG GND};
      \draw (25,2)--(25,0)to[C=$C_{13}$,-*](27,0);
      \draw (27,5)node[anchor=south]{Anode}to[short,*-o](28,5) node[anchor=west] {SIG OUT};
      \draw (27,2)to[short] (27,-1) to[R=$R_i$] (27,-4) node[ground]{GND}; %
      \draw (25,-2) node[align=left,text width=20cm, scale=2] 
  {$R_i$:1k$\Omega$,  R1-R3:360k$\Omega$,  R4-R13:330k$\Omega$, C1-C13:10nF  };

    \end{circuitikz} 
    \end{centering}
    \caption{Schematic diagram of the secondary emission voltage divider for R5900-00-M16 PMTs.}
    \label{SE Diagram for R5900}
\end{figure}
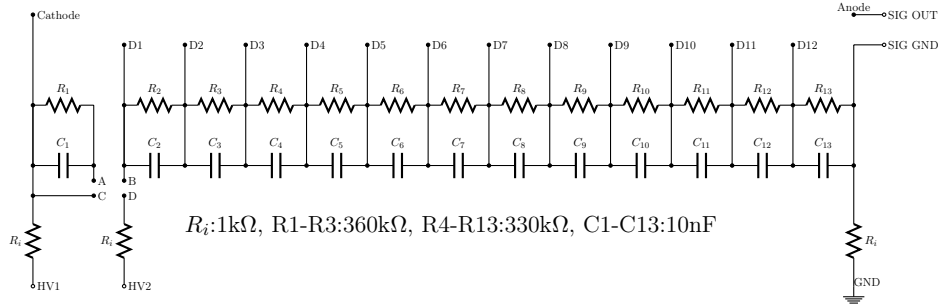

There are only two different modes of operation for the baseboards of R5900-00-M16 multi-anode PMTs: Mode 1 and Mode 2. The schematic diagram of the secondary emission voltage divider for R5900-00-M16 is shown in Fig. \ref{SE Diagram for R5900}. The A-B bridge forms Mode 1 and the B-C bridge forms Mode 2, both with HV input on HV1.

\begin{figure}[h!]
  \begin{centering}
    \begin{circuitikz}[xscale=0.34, yscale=0.4, transform shape]
      \draw (0,-3)node[anchor=west] {HV1} to[R=$R_i$,o-] (0,0) to[short,-*] (0,5) node[anchor=west] {Cathode}; 
      \draw (0,2)to[R=$R_1$,*-] (2,2) to[short,-*] (2,0.5)node[anchor=west] {A}; 
      \draw (0,0)to[short,*-*](2,0) node[anchor=west]{C};
      \draw (3,-3) node[anchor=west] {HV2} to[R=$R_i$,o-*] (3,0)node[anchor=west] {D};
      \draw (3,0.5) node[anchor=west] {B}to[short,*-*] (3,4) node[anchor=west] {D1};
      \draw (3,2)to[R=$R_2$,*-*] (5,2) to[short,-*](5,4) node[anchor=west] {D2};
      \draw (5,2)to[R=$R_3$,-*] (7,2) to[short,-*](7,4) node[anchor=west] {D3}; 
      \draw (7,2)to[R=$R_4$,-*] (9,2) to[short,-*](9,4) node[anchor=west] {D4};
      \draw (9,2)to[R=$R_5$,-*] (11,2) to[short,-*](11,4) node[anchor=west] {D5};
      \draw (11,2)to[R=$R_6$,-*] (13,2) to[short,-*](13,4) node[anchor=west] {D6};
      \draw (13,2)to[R=$R_7$,-*] (15,2) to[short,-*](15,4) node[anchor=west] {D7};
      \draw (15,2)to[R=$R_8$,-*] (17,2) to[short,-*](17,4) node[anchor=west] {D8};
      \draw (17,2)to[R=$R_9$,-*] (19,2) to[short,-*](19,4) node[anchor=west] {D9};
      \draw (19,2)to[R=$R_{10}$,-*] (21,2) to[short,-*](21,4) node[anchor=west] {D10};
      \draw (21,2)to[R=$R_{11}$,-*] (23,2) to[short,-*](23,4) node[anchor=west] {D11};
      \draw (23,2)to[R=$R_{12}$,-*] (25,2) to[short,-*](25,4) node[anchor=west] {D12};
      \draw (25,2)to[R=$R_{13}$,-*] (27,2) to[short,-*](27,4) node[anchor=west] {D13};
      \draw (27,2)to[R=$R_{14}$,-*] (29,2) to[short,-*](29,4) node[anchor=west] {D14};
      \draw (29,2)to[R=$R_{15}$,-*] (31,2) to[short,-*](31,4) node[anchor=west] {D15};
      \draw (31,2)to[R=$R_{16}$,-*] (33,2) to[short,-*](33,4) node[anchor=west] {D16};
      \draw (31,2)--(31,0)to[C=$C_1$,-*](33,0);
      \draw (33,2)to[R=$R_{17}$,-*] (35,2) to[short,-*](35,4) node[anchor=west] {D17};
      \draw (33,2)--(33,0)to[C=$C_2$,-*](35,0);
      \draw (35,2)to[R=$R_{18}$,-*] (37,2) to[short,-*](37,4) node[anchor=west] {D18};
      \draw (35,2)--(35,0)to[C=$C_3$,-*](37,0);
      \draw (37,2)to[R=$R_{19}$,-*] (39,2) to[short,-*](39,4) node[anchor=west] {D19};
      \draw (37,2)--(37,0)to[C=$C_4$,-*](39,0);
      \draw (39,2)to[R=$R_{20}$,-*] (41,2) to[short](41,4) to[short,-o] (42,4)node[anchor=west] {SIG GND};
      \draw (39,2)--(39,0)to[C=$C_5$,-*](41,0);
      \draw (41,5)node[anchor=south]{Anode}to[short,*-o](42,5) node[anchor=west] {SIG OUT};
      \draw (41,2)to[short] (41,0) to[R=$R_i$] (41,-2) node[ground]{GND}; %
      \draw (25,-1) node[align=left,text width=20cm, scale=2] 
  {$R_i$:1k$\Omega$,  R1:660k$\Omega$,  R2-20:330k$\Omega$, C1-5:10nF  };
    \end{circuitikz}
    \caption{Schematic diagram of the secondary emission baseboard common voltage divider for R7761 PMTs.}
    \label{SE Diagram for R7761}
  \end{centering}
\end{figure}
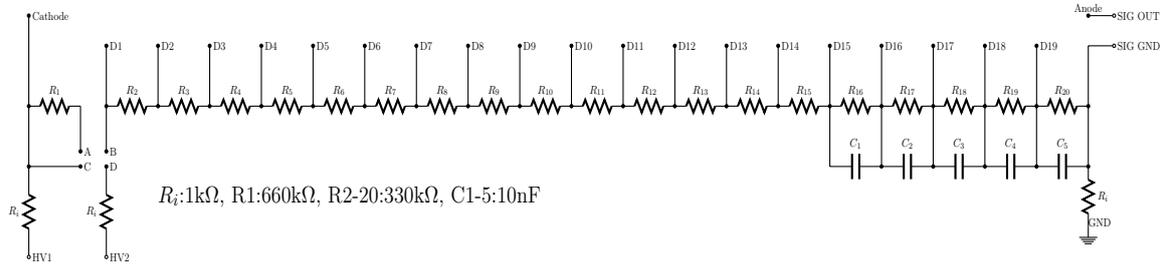

\section{Measurements and Results}

The PM and SE mode characterization tests were done at the University of Iowa PMT test station \citep{Akgun2014,Tiras2012}. All the tests were performed in light-tight boxes. Each anode of the R5900-00-M16 multi-anode PMT was tested for response uniformity. A 20 Hz, 337 nm nitrogen laser was used as a source with a neutral density filter and the light was transmitted to the window of the PMT by a 1 mm diameter wavelength shifting optical fiber. The fiber was placed at the center of each anode and the measurements were taken across all 16 anodes.  

\begin{figure}[]
\centering
\includegraphics[scale=0.50]{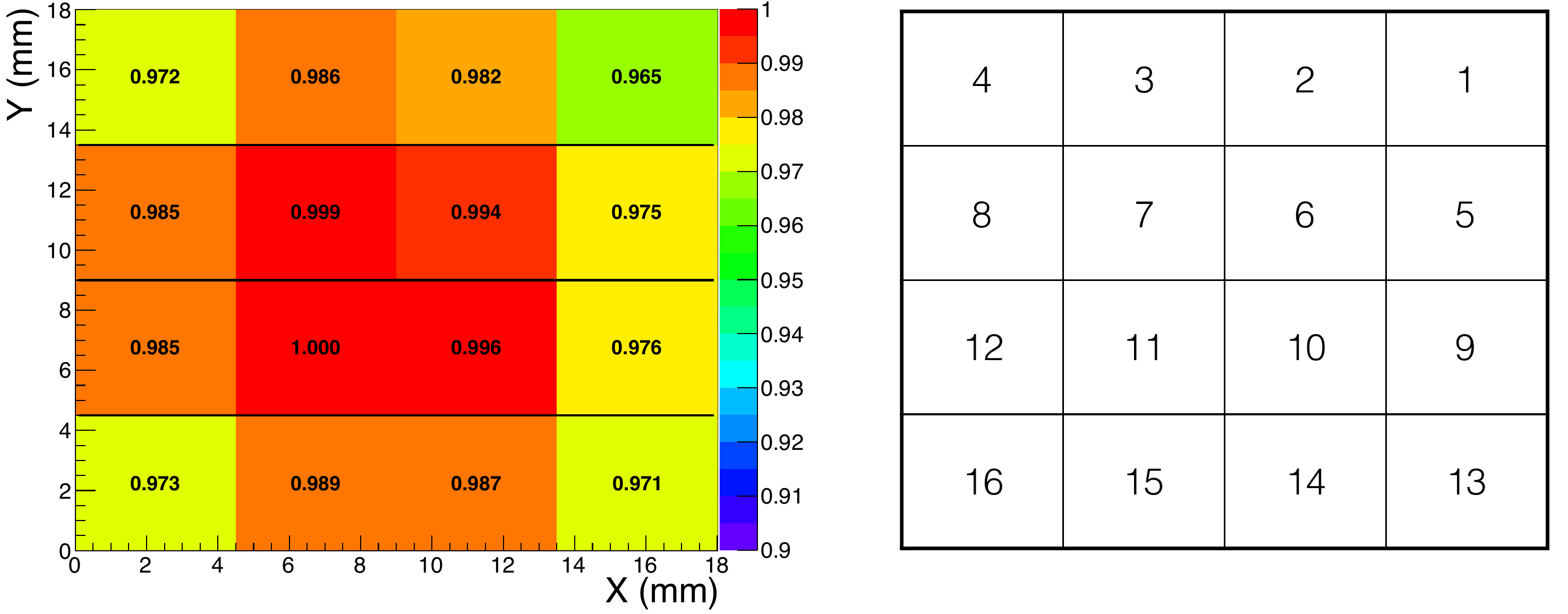}
\caption{The response uniformity of R5900-00-M16 PMT (left) as light incidence on channels 1-16, respectively. Right is the top view of R5900-00-M16 PMT, the integers indicate the channel numbers.}
\label{Nonuniformity} 
\end{figure}

Figure \ref{Nonuniformity} shows the response uniformity measurement for a R5900-00-M16 PMT. The responses were normalized to that of anode 11. The non-uniformity is measured to be less than 3\%. This result shows that the variation of sensitivity across the effective area of multianode PMTs is low enough to be considered as active layers in calorimetric measurements with beams. 

For the dark current measurements, each PMT was placed in a light-tight box, and the anode current is read out by the Keithley-6485 picoammeter right after the current was settled. The dark current distributions for both single anode R7761 PMTs and multi-anode R5900-00-M16 PMTs show that the majority of the PMTs have dark currents below 1 nA at the range of operation voltages 1500-2000 V and 800-1200 V respectively, for all the different operation modes. 

The anode and cathode currents at various light intensities were measured in a light tight box and gains were calculated for both PMT types in all modes. The purpose of the measurements is to obtain a relative gain for the SE modes, i.e. for multiplication starting at D1 instead of the photocathode, in comparison with the conventional PMT mode. In these tests, uniformly distributed light pulses are used to illuminate the PMT window. A tungsten light in the opposite end of the light-tight box is used as a light source. A neutral density filter, which reduces the light intensity to the desired level, and a light diffuser, which distributes the light uniformly through the face of the PMT, are used in the upstream direction of the PMT. Since there is no natural way of initiating a multiplication chain such as thermal excitation, we tried to forcefully generate a photoelectron from the photocathode with a light source and have it multiplied with the modified bias arrangement. 

\begin{figure}[H]
\centering
\includegraphics[width=0.45\textwidth]{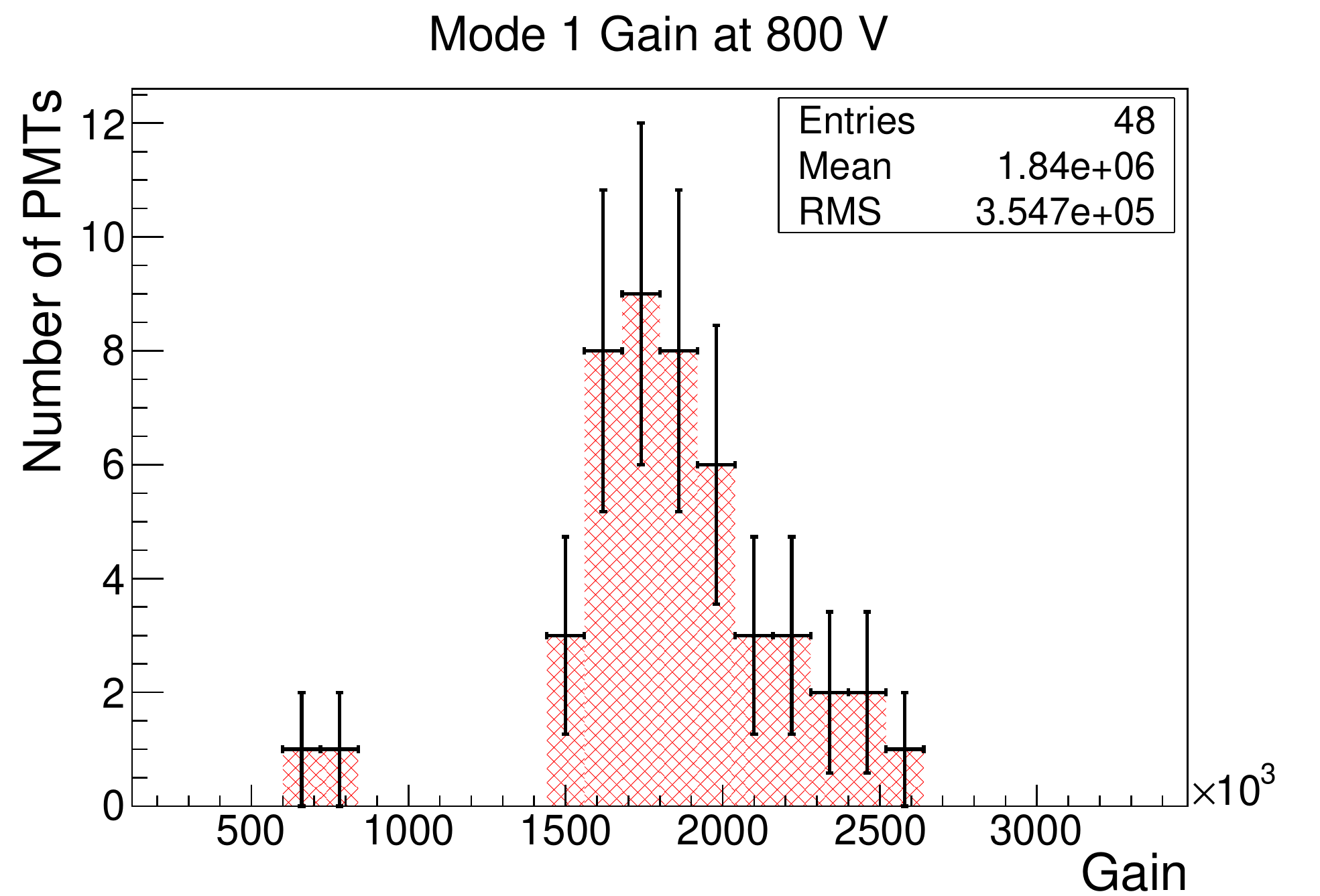}
\includegraphics[width=0.45\textwidth]{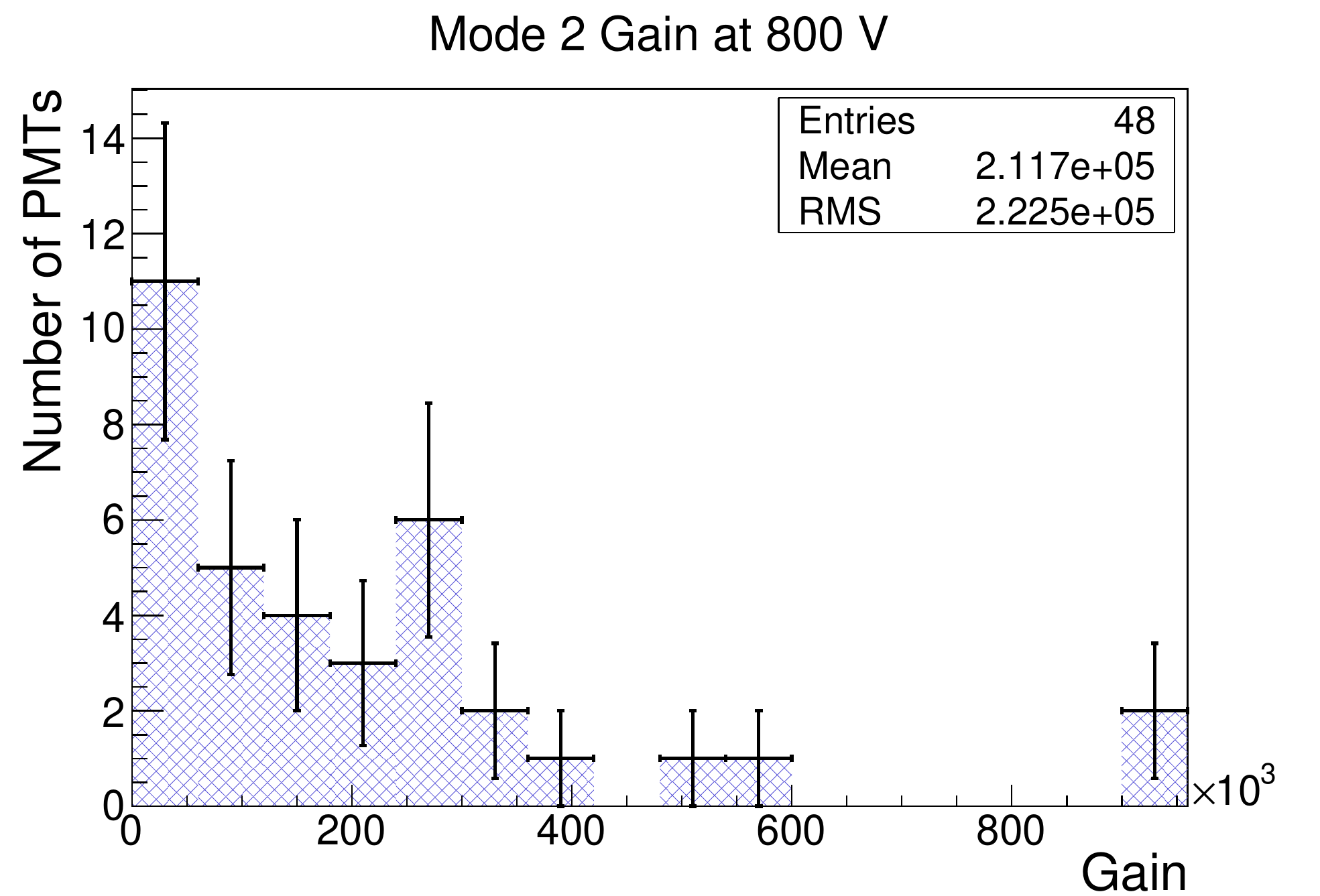}
\includegraphics[width=0.45\textwidth]{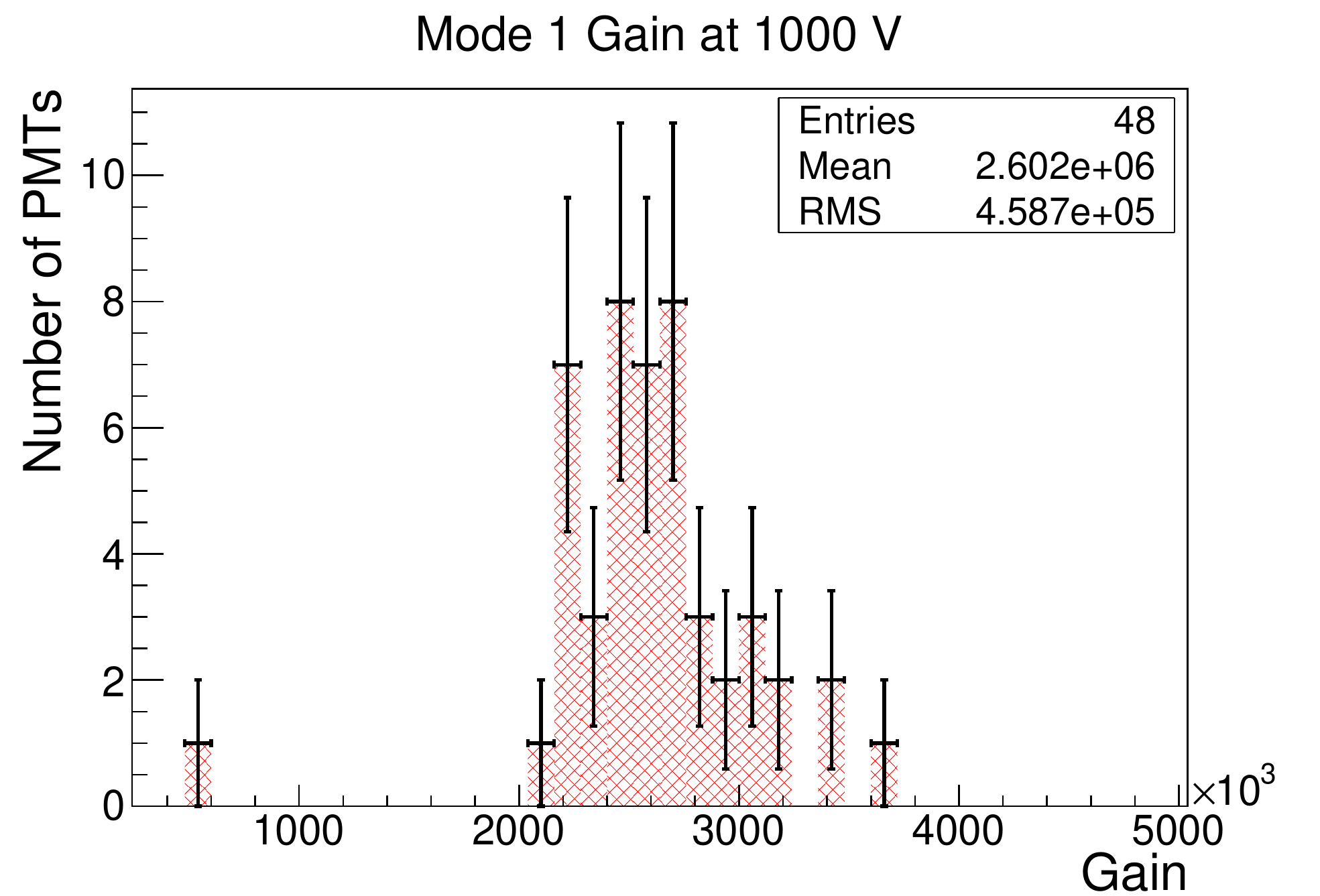}
\includegraphics[width=0.45\textwidth]{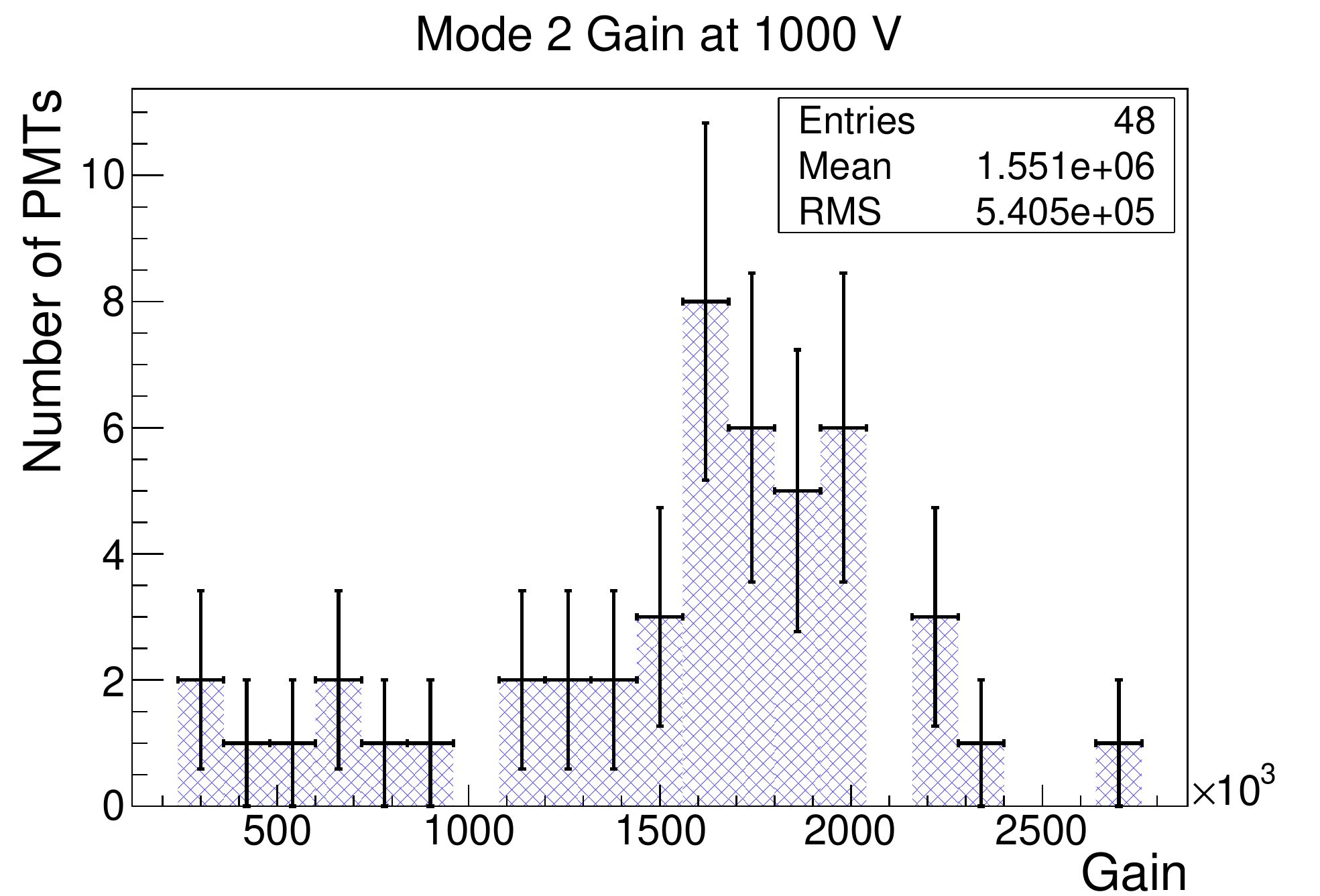}
\includegraphics[width=0.45\textwidth]{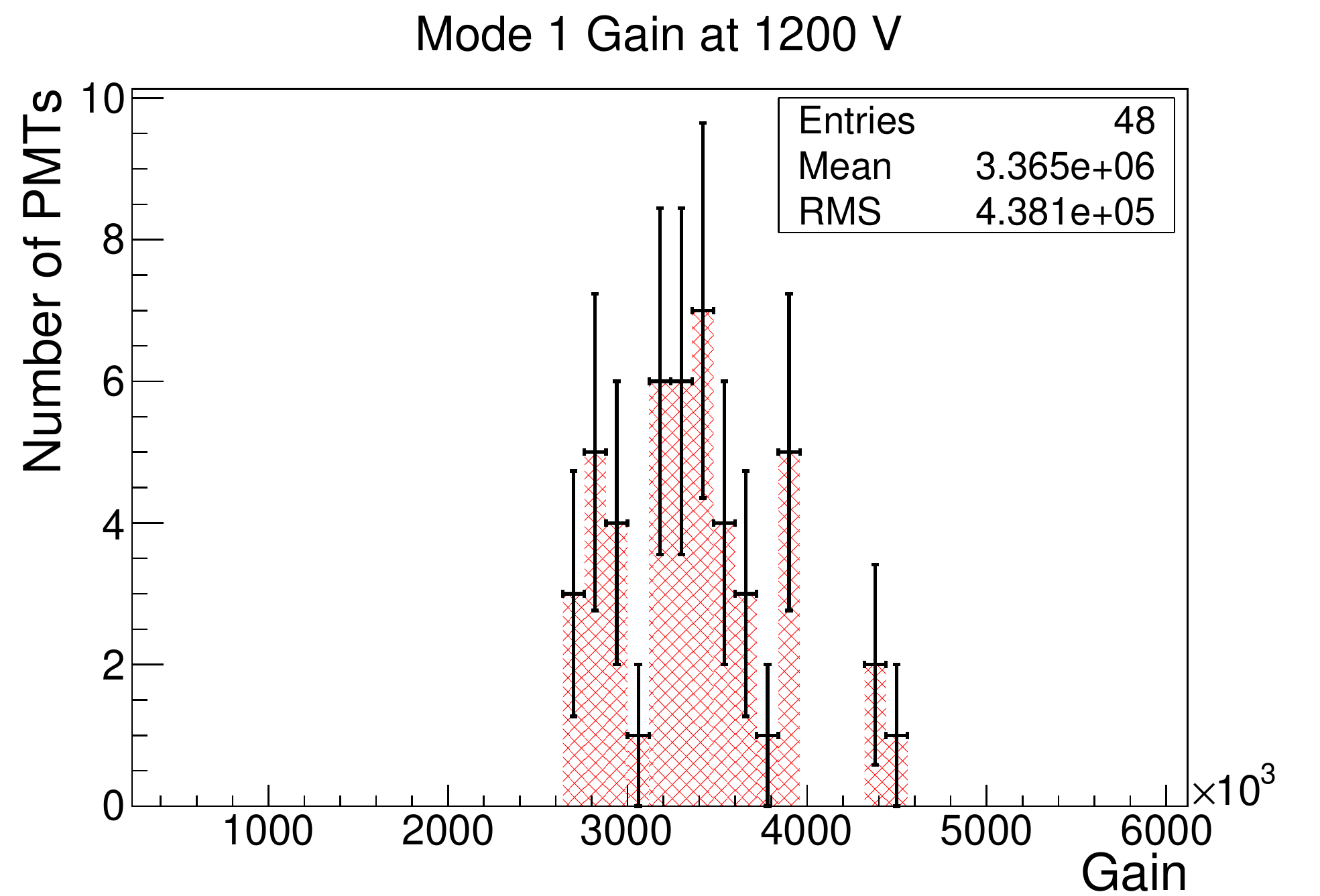}
\includegraphics[width=0.45\textwidth]{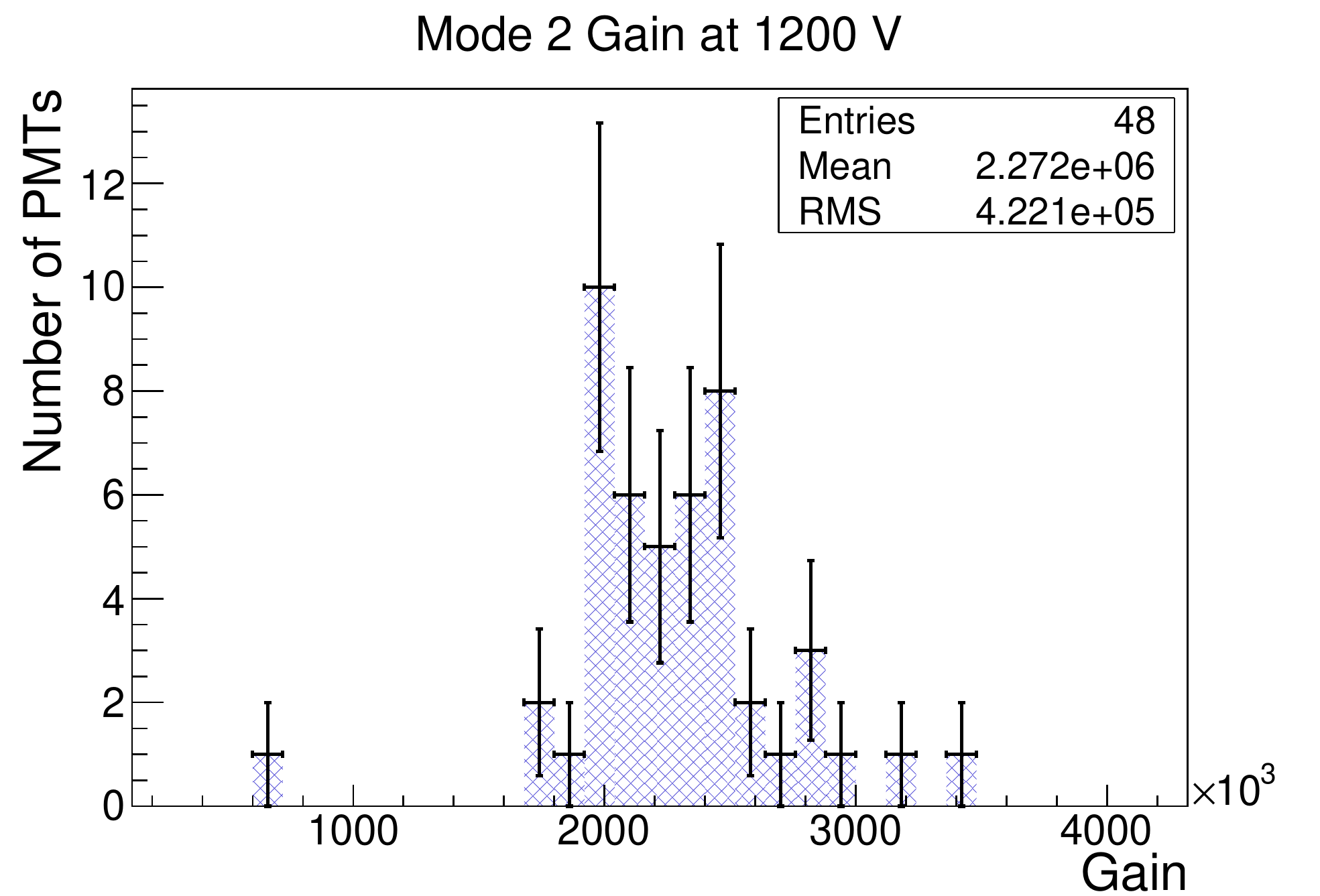}
\caption{Gain distributions of Mode 1 (left column) and Mode 2 (right column) for R5900-00-M16 PMTs at 800 V (top row), 1000 V (middle row) and 1200 V (bottom row).}
\label{GainR5900M16} 
\end{figure}

Figure \ref{GainR5900M16} shows the gain distributions of R5900-00-M16 PMTs in Mode 1 and Mode 2 at 800, 1000 and 1200 V. This is the total gain of all 16 anodes. Table \ref{5900 Gain Summary} summarizes Mode 1 and Mode 2 gains (Mean and RMS). The performance of Mode 2 at 1000 V with and average gain of 16$\times 10^{5}$ is comparable to the gain of Mode 1 at 800 V. The decrease in the gain due to the smaller number of steps in the multiplication chain can be roughly compensated with an increase of 200 V in the operating voltage. As the gain of Mode 2 rises to the Mode 1 levels, the spread of the gains across different Mode 2 PMTs gets smaller. This is an indication that with the combination of the stable, dedicated SE modules and the correct operating conditions, large sizes of calorimeters based on SE principle can be constructed. 

\begin{figure}[H]
\centering
\includegraphics[width=0.32\textwidth]{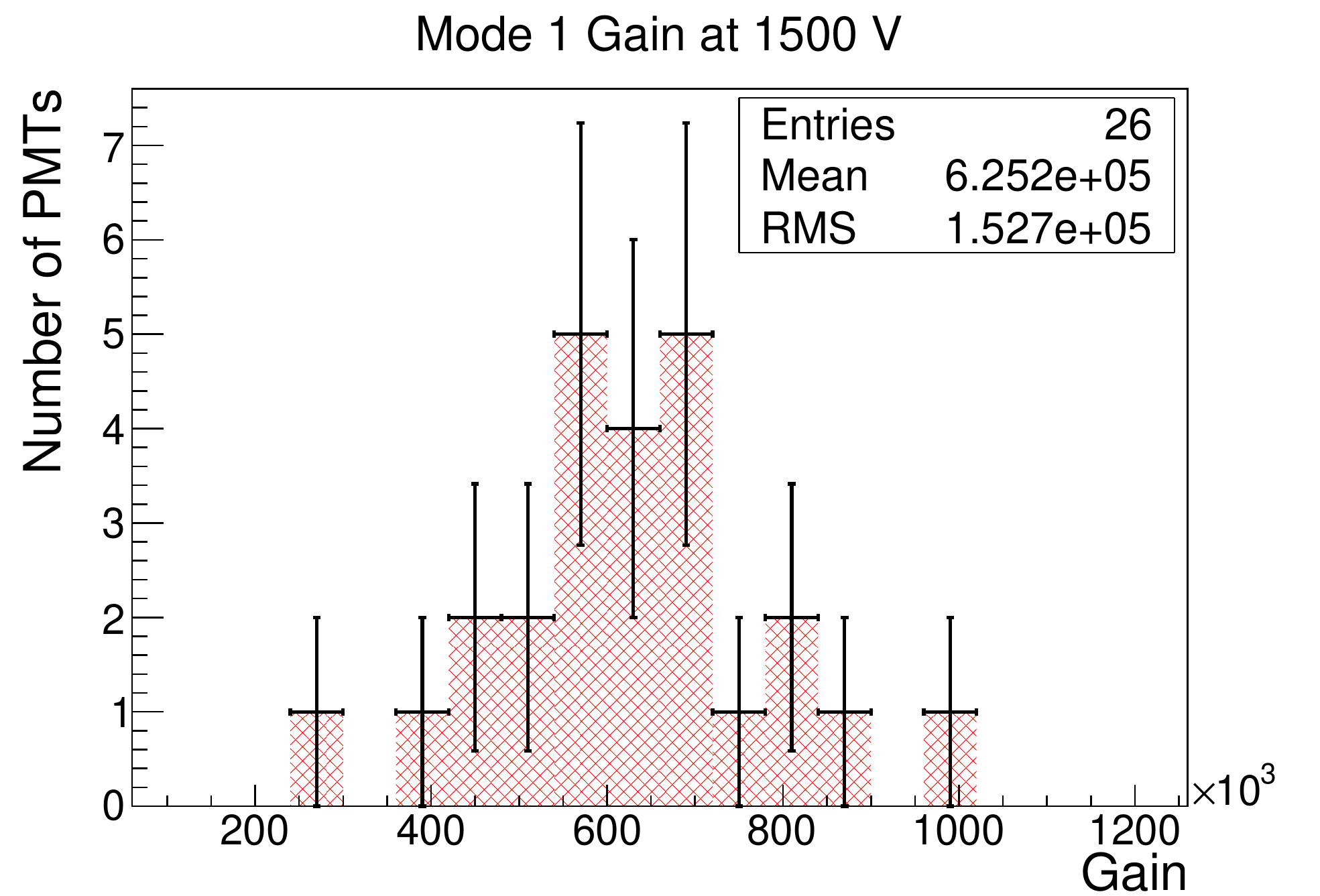}
\includegraphics[width=0.32\textwidth]{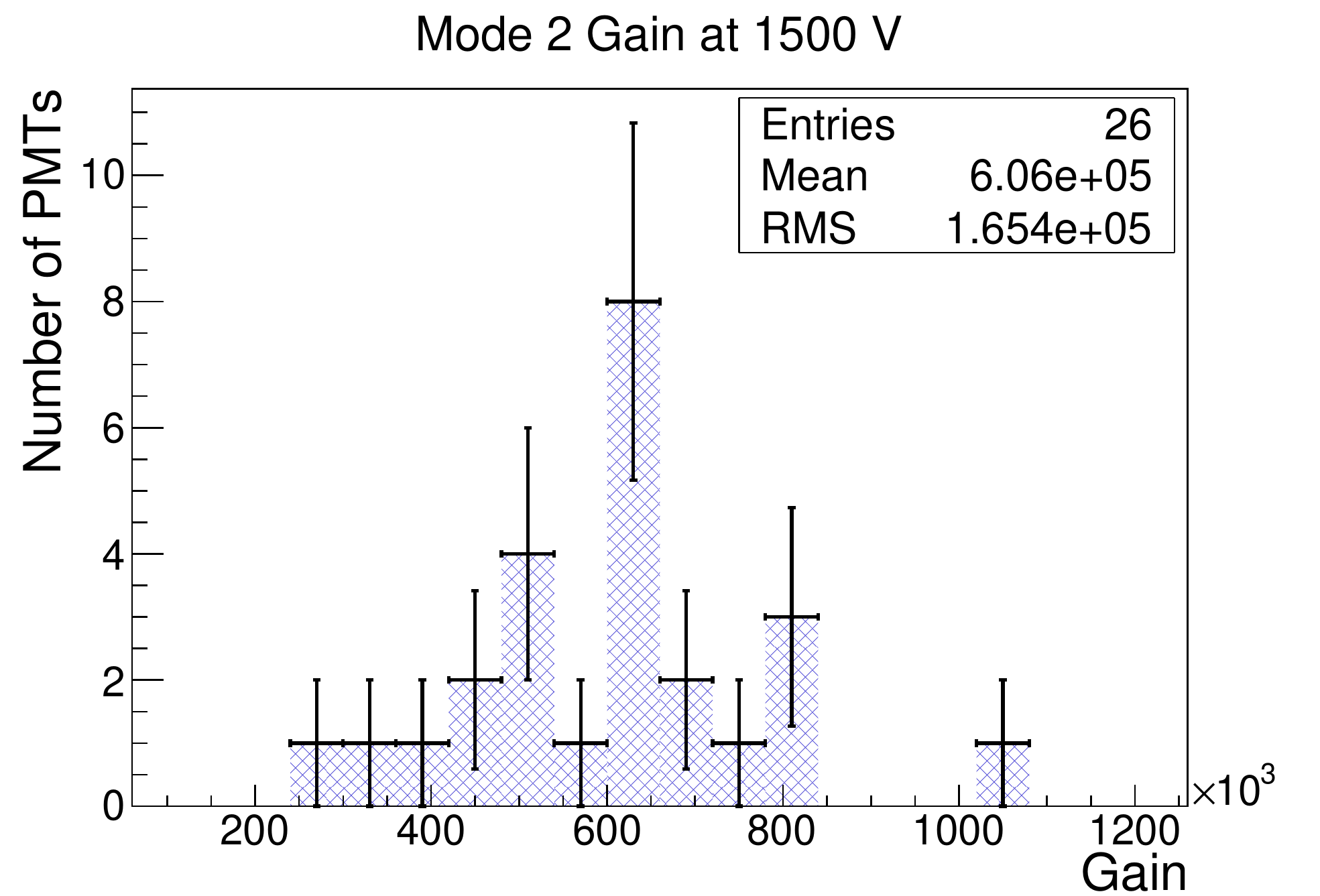}
\includegraphics[width=0.32\textwidth]{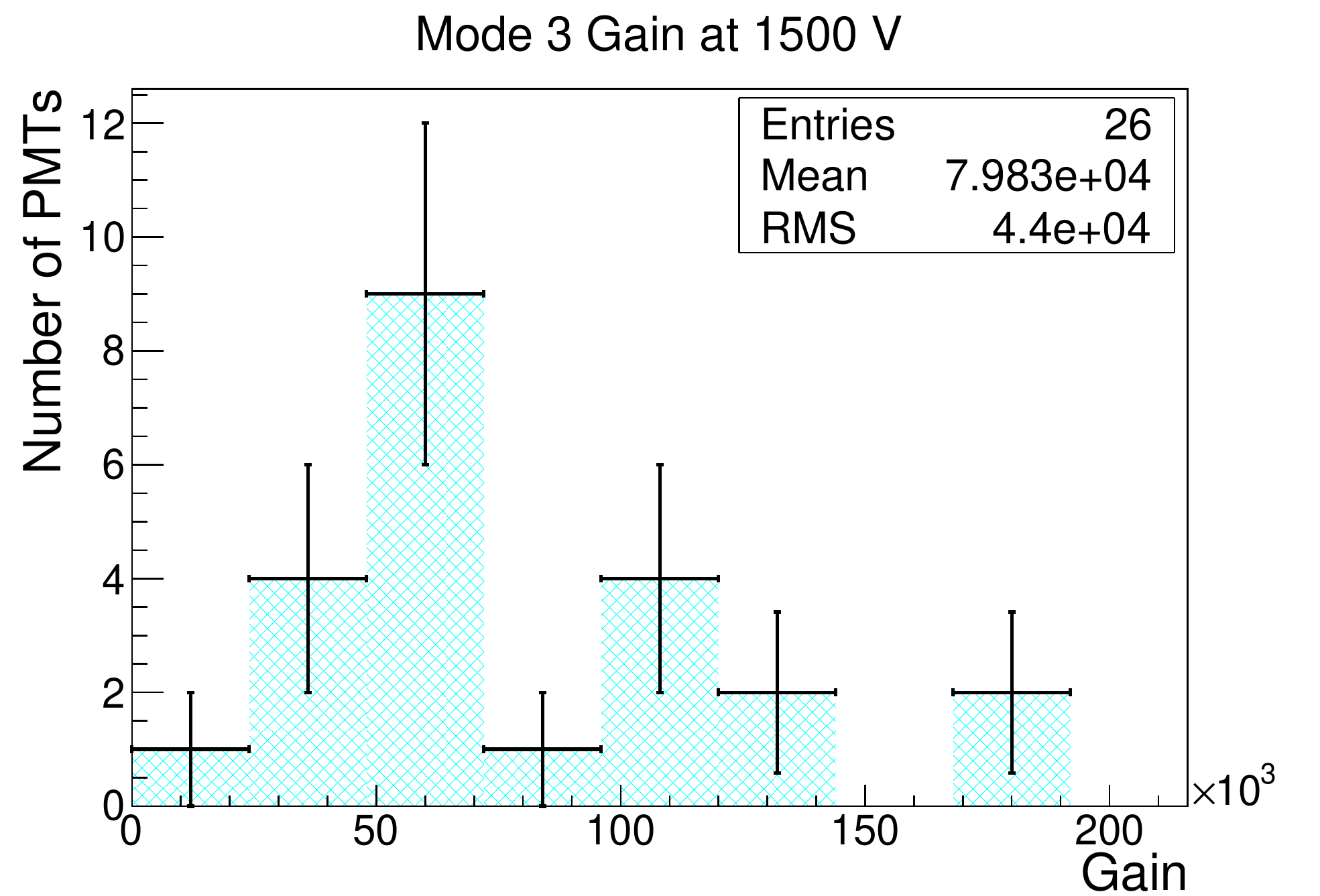}
\includegraphics[width=0.32\textwidth]{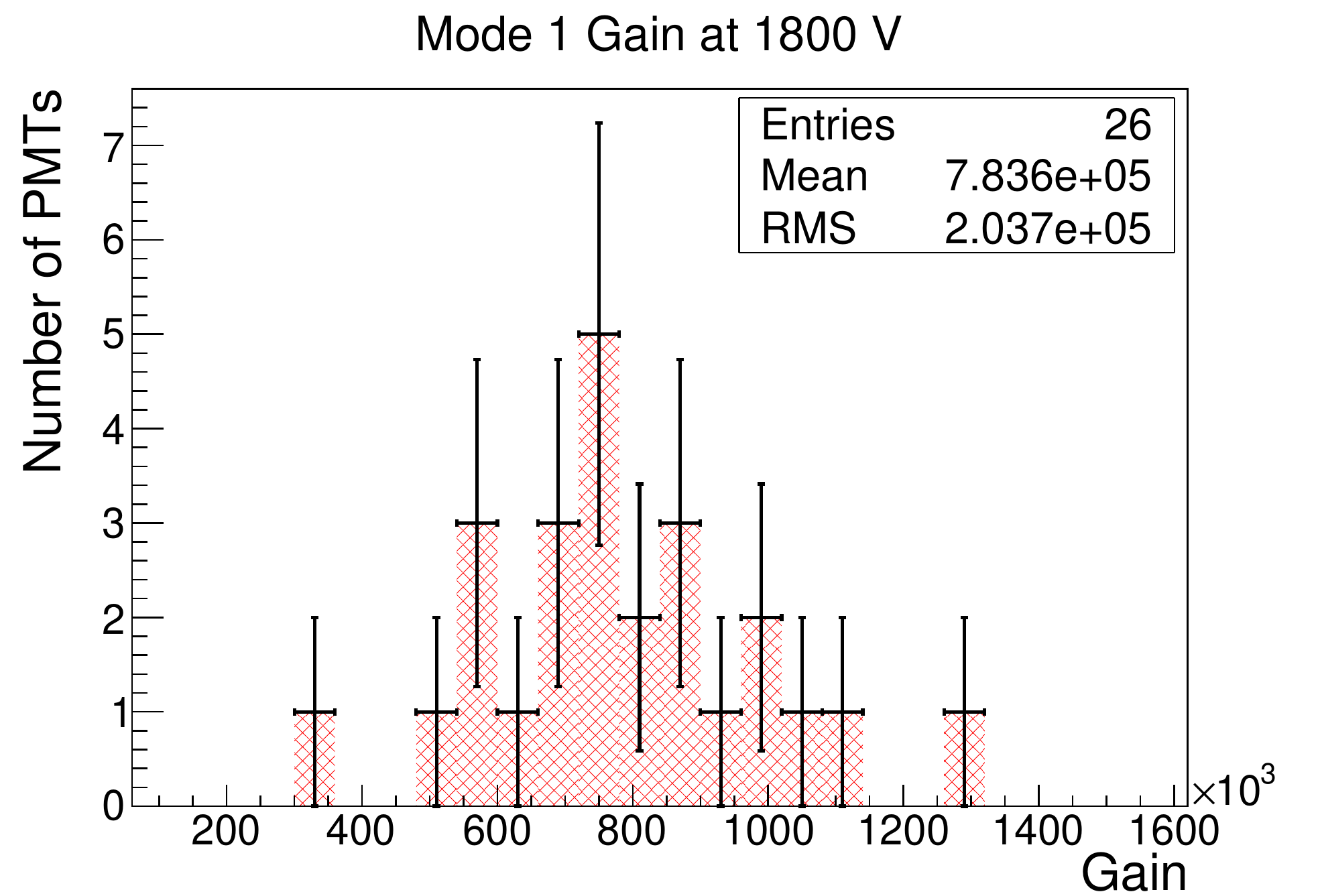}
\includegraphics[width=0.32\textwidth]{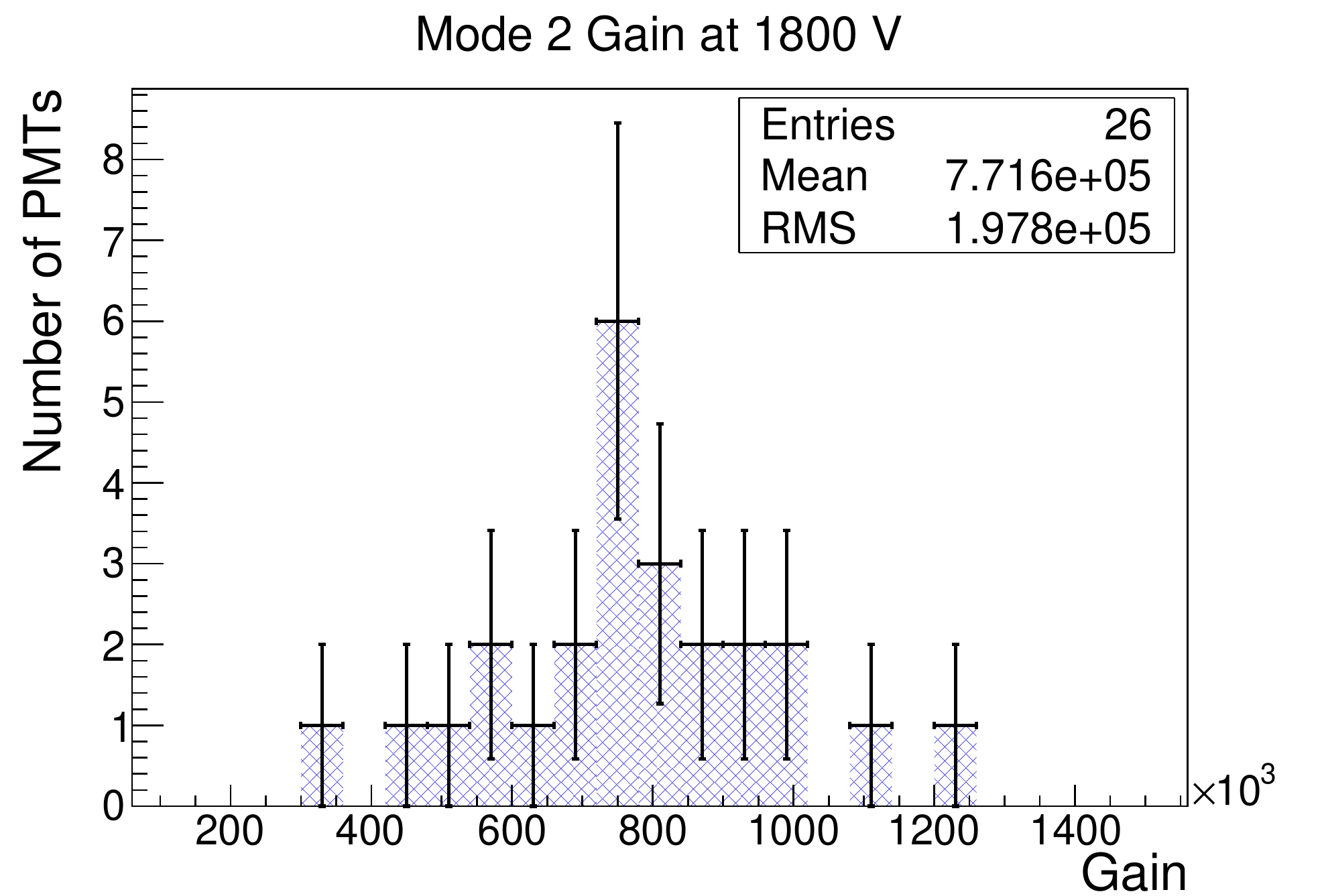}
\includegraphics[width=0.32\textwidth]{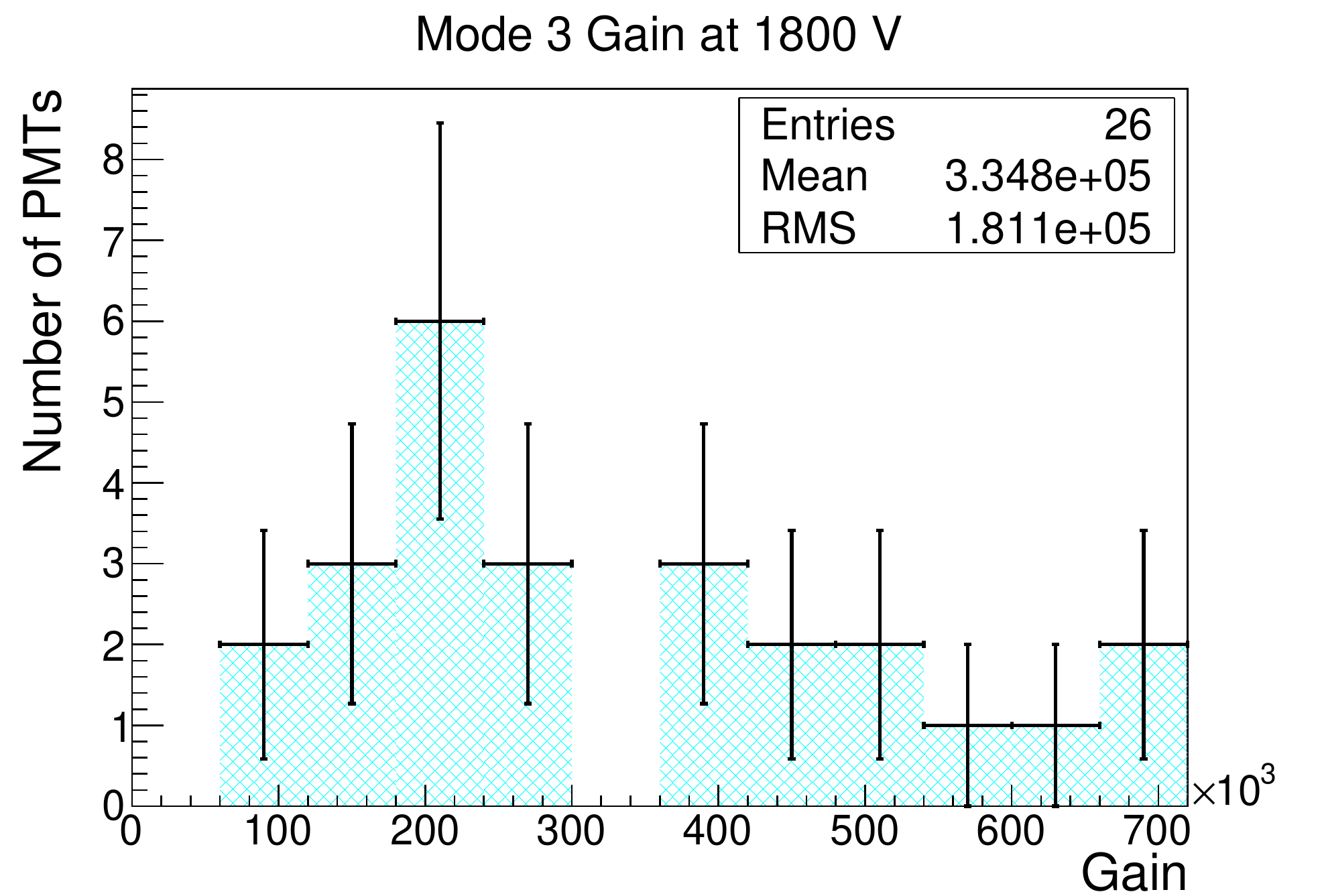}
\includegraphics[width=0.32\textwidth]{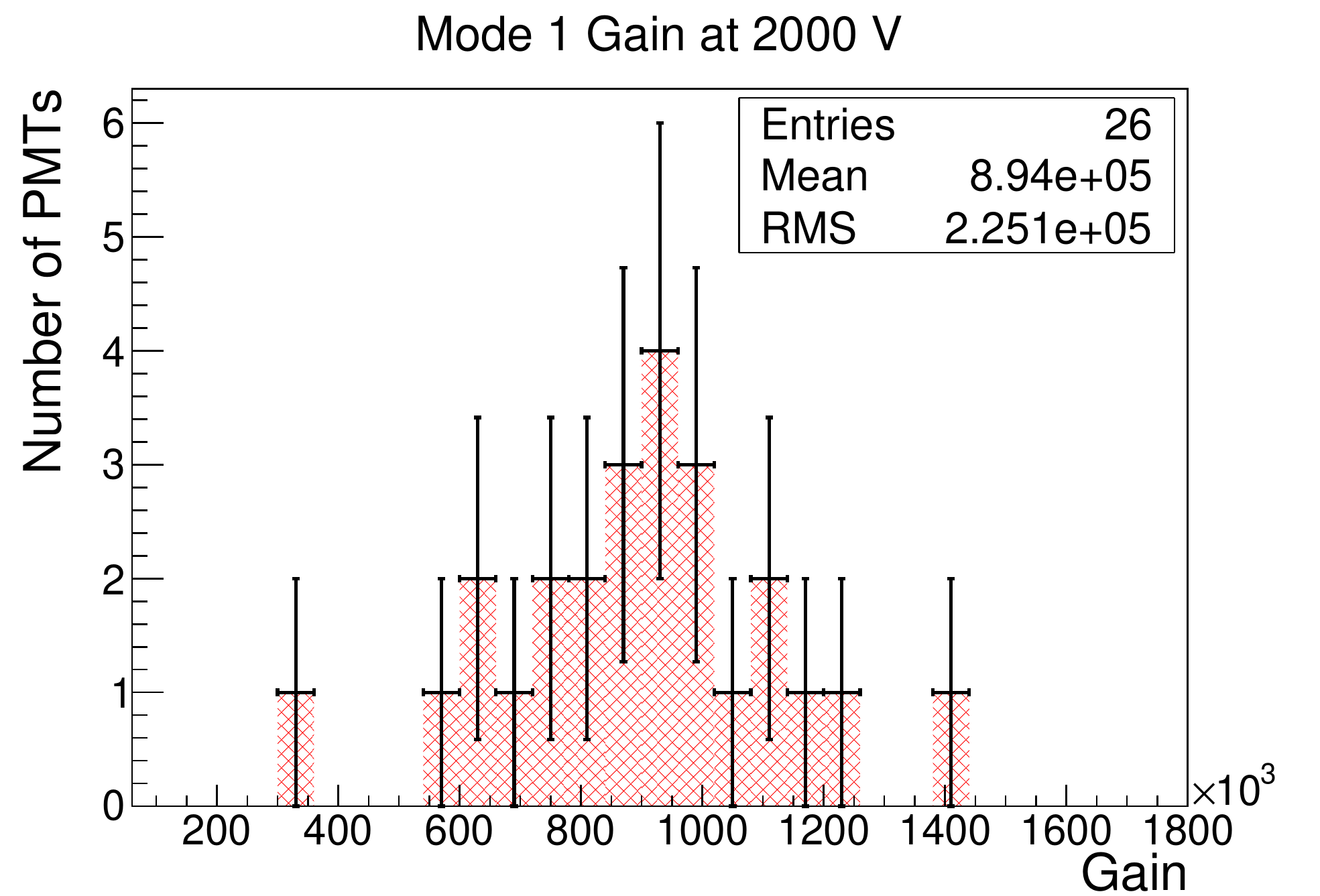}
\includegraphics[width=0.32\textwidth]{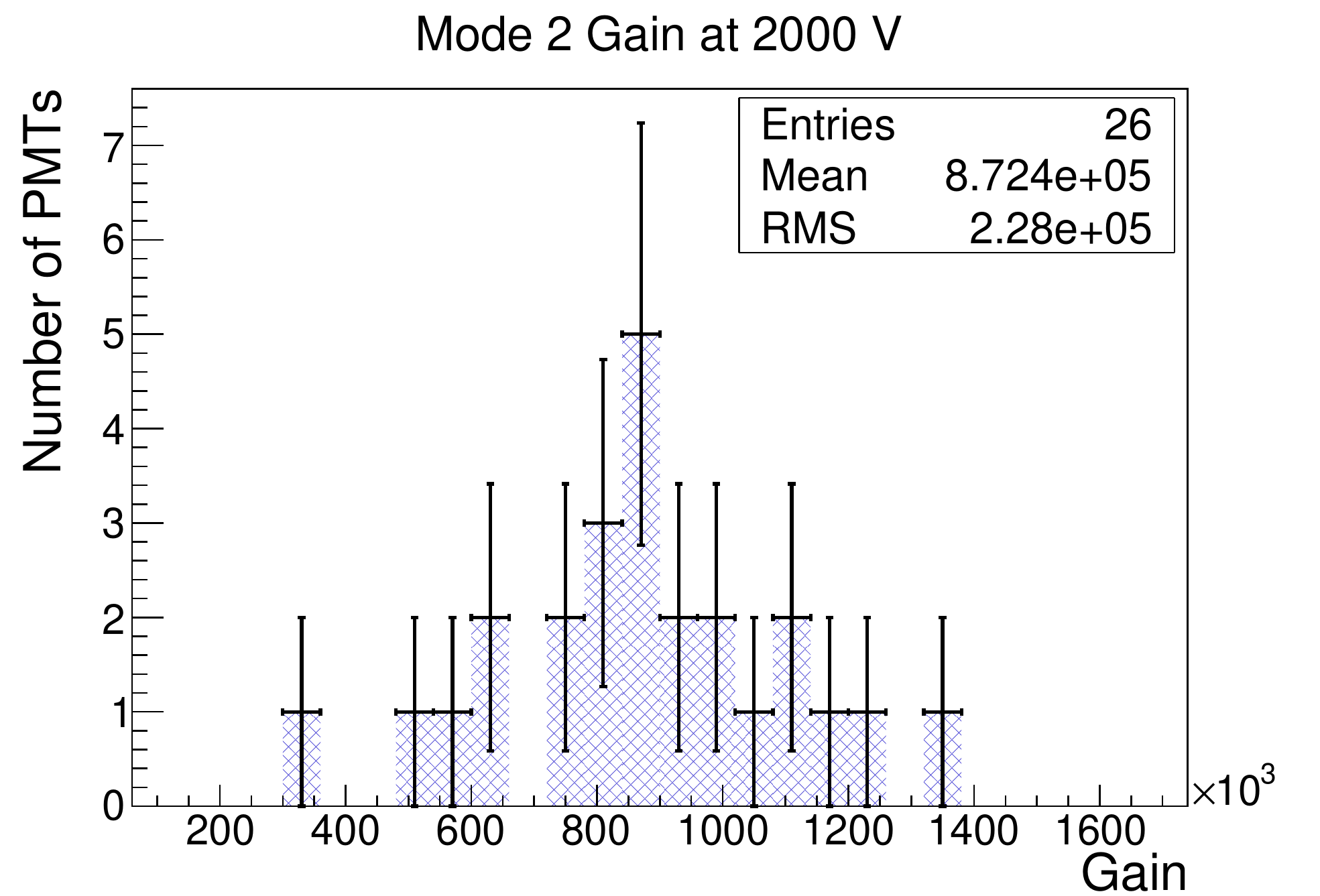}
\includegraphics[width=0.32\textwidth]{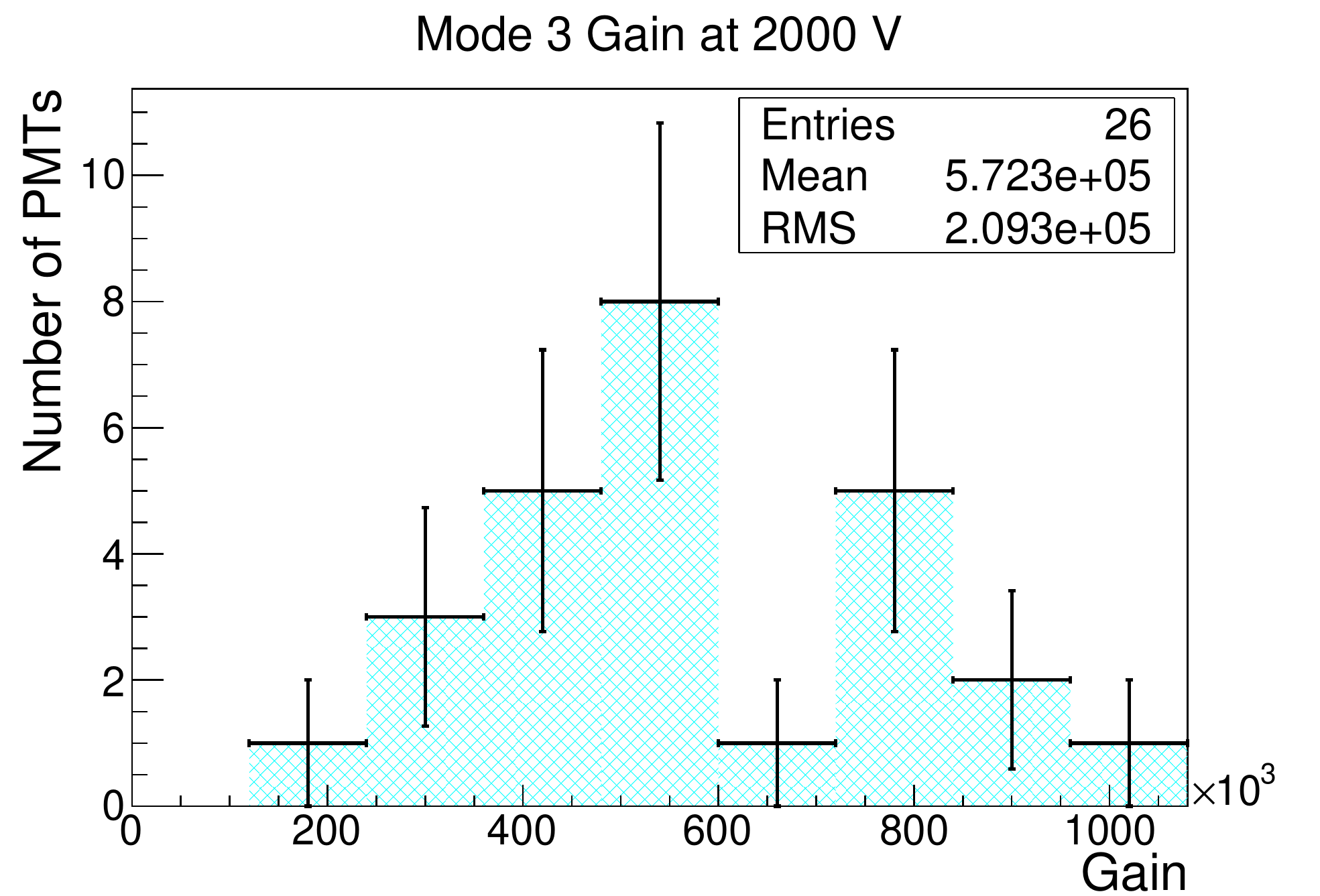}
\caption{Gain distributions of Mode 1 (leftmost column), Mode 2 (middle column) and Mode 3 (rightmost column) for R7761 PMTs at 1500 V (top row), 1800 V (middle row) and 2000 V (bottom row).}
\label{GainR7761} 
\end{figure}

Figure \ref{GainR7761} shows the gain distributions of R7761 PMTs in Mode 1, Mode 2 and Mode 3 at 1500, 1800 and 2000 V. Table \ref{R7761 Gain Summary} summarizes gains (Mean and RMS) of all mentioned modes. The performance of Mode 2 is almost identical to Mode 1 at the range of operation voltages 1500 - 2000 V. The advantage of the mesh dynode structure over the metal channel structure is clearly visible. The geometry and size of the metal channels further constrain the favorable location of the first SEe, hence a larger voltage difference is needed to compensate for this effect. On the other hand, the mesh dynodes constitute a more uniform surface (like the photocathode). Therefore, the gains in Mode 2 are comparable to Mode 1 in contrast with R5900-00-M16 PMTs. 

The significantly lower gain of Mode 3 is understood to be due to some of the SEe not being able to reach D2 and be amplified in the dynode chain. Here, the relative electric field between the floating cathode and D1, and between D1 and D2 play an important role. Even small fluctuations seem to yield lower gains. The performance of Mode 3 at 2000 V with a gain of 5.72$\times 10^{5}$ is hardly comparable to those of Mode 1 and Mode 2 at 1500 V.\raggedbottom

The statistical uncertainty of gains and dark currents were determined after retesting the same PMTs several times and it was about 10\% for both gain and dark current measurements.

\begin{table}[H]
\caption{Summary of gains for Mode 1 and Mode 2 of R5900-00-M16 PMTs.}
\label{5900 Gain Summary}
\begin{center}
\begin{tabular}{lcccccc}
\toprule 
\multirow{2}{*}{Gains ($\times 10^{5}$)} & 
\multicolumn{2}{c}{800 V} &
\multicolumn{2}{c}{1000 V} &
\multicolumn{2}{c}{1200 V} \\
      & {Mean} & {RMS} & {Mean} & {RMS} & {Mean} & {RMS} \\
      \midrule
    Mode 1 & 18.4 & 3.55 & 26.02 & 4.59 & 33.65 & 4.38 \\
    Mode 2 & 2.12 & 2.23 & 15.51 & 5.41 & 22.72 & 4.22 \\
    \bottomrule
\end{tabular}
\end{center}
\end{table}

\begin{table}[H]
\caption{Summary of gains for Mode 1, Mode 2 and Mode 3 of R7761 PMTs}
\label{R7761 Gain Summary}
\begin{center}
\begin{tabular}{lcccccc}
\toprule 
\multirow{2}{*}{Gains ($\times 10^{5}$)} & 
\multicolumn{2}{c}{1500 V} &
\multicolumn{2}{c}{1800 V} &
\multicolumn{2}{c}{2000 V} \\
      & {Mean} & {RMS} & {Mean} & {RMS} & {Mean} & {RMS} \\
      \midrule
    Mode 1 & 6.25 & 1.53 & 7.84 & 2.04 & 8.94 & 2.25 \\
    Mode 2 & 6.06 & 1.65 & 7.72 & 1.98 & 8.72 & 2.28 \\
    Mode 3 & 0.80 & 0.44 & 3.35 & 1.81 & 5.72 & 2.09 \\
    \bottomrule
\end{tabular}
\end{center}
\end{table}

\section{Conclusions}

We developed secondary emission readout modes as a simple bias modification in the voltage dividers of photomultiplier tubes with mesh and metal channel dynode structures. In the first secondary emission module, seven single anode Hamamatsu R7761 PMTs were powered by a dedicated circuit board, which can be operated in three diffrent modes; normal divider (Mode 1), cathode-first dynode shorted (Mode 2) and cathode float mode (Mode 3). The second secondary emission module was produced with nine multianode Hamamatsu R5900-00-M16 PMTs, powered by separate bases, which can be used in two different modes; normal divider (Mode 1) and cathode-first dynode shorted mode (Mode 2). Both PMT types run smoothly in both traditional PM and SE modes, even after being used to collect data at the CDF experiment at Fermilab for several years. All the PMTs have low dark currents, below 1 nA for all the different operation modes. The multianode PMTs have a response non-uniformity of less than 3\% across the anodes. Average gain for the R5900-00-M16 in Mode 1 is 15-30$\times10^5$ and in Mode 2 is 2-20$\times10^5$ for the range of operating voltages 800-1200 V. Average gain for R7761 is 6-9$\times10^5$ both for Mode 1 and Mode 2, and in Mode 3 is 1-6$\times10^5$ for the range of operating voltages 1500-2000 V. Both modules are operating stably and can serve as active layers in calorimetric measurements with beams in order to validate the concept of secondary emission calorimetry. 

\acknowledgments

The authors would like to thank CDF Collaboration for loan of PMTs.


\begin{thebibliography}{1}

\bibitem{PhysRev.107.977}
Robert~G. Lye and A.~J. Dekker, 
\newblock Theory of secondary emission.
\newblock {\em Phys. Rev.}, 107:977--981, Aug 1957.

\bibitem{winnonel}
David R. Winn and Yasar Onel,
\newblock{Secondary Emission Calorimeter Sensor Development} 
\newblock{\em Journal of Physics: Conference Series}, V404-1-012021, 2012, 
\url{http://iopscience.iop.org/article/10.1088/1742-6596/404/1/012021}. 

\bibitem{snowmass}
A. Albayrak-Yetkin and et~al., 
\newblock{Secondary Emission Calorimetry: Fast and Radiation Hard},  
\newblock{\em Community Summer Study 2013: Snowmass on the Mississippi (CSS2013) Minneapolis}, arXiv:1307.8051, 2013. 
\url{https://inspirehep.net/record/1245371/files/arXiv:1307.8051.pdf}.

\bibitem{CMSHCAL}
CMS HCAL Collaboration,
\newblock {Study of various photomultiplier tubes with muon beams and Cerenkov light produced in electron showers}.
\newblock {\em Journal of Instrumentation},V5-T06002, June 2010.

\bibitem{R7761}
Hamamatsu Photonics, 2016, 
\newblock
  \url{https://www.hamamatsu.com/resources/pdf/etd/High_energy_PMT_TPMO0007E.pdf}. 

\bibitem{Artikov}
A.~Artikov and et~al.,
\newblock {CDF central preshower and crack detector upgrade}.
\newblock FNAL-PUB-07-023-E, 2007, 
\url {https://arxiv.org/pdf/0706.3922.pdf}. 


\bibitem{Grozis}
C.~Grozis and et~al.,
\newblock {The time-of-flight detector at CDF}.
\newblock {\em Nucl. Phys. Proc. Suppl.}, 93:344--347, 2001.

\bibitem{Blair}
The CDF II Collaboration.,
\newblock {The CDF II Detector Technical Design Report}, FERMILAB Pub-96/390-E,
\newblock 1996, 
\url {http://lss.fnal.gov/archive/design/fermilab-design-1996-01.pdf}.

\bibitem{Akgun2014}
U.~Akgun and et~al.,
\newblock {Characterization of 1800 Hamamatsu R7600-M4 PMTs for CMS HF
  Calorimeter upgrade}.
\newblock {\em Journal of Instrumentation},V9-T06005, June 2014.

\bibitem{Tiras2012}
E.~Tiras.,
\newblock {Characterization of 900 four-anode photomultiplier tubes for use in 2013 hadronic forward calorimeter upgrade}.
\newblock {M.Sc.}, University of Iowa, 2012, 
\url{http://ir.uiowa.edu/etd/3392}. 

\end{thebibliography}

\end{document}